%Paper: hep-ph/9310290
%From: Thomas <thomas@scipp.ucsc.edu>
%Date: Thu, 14 Oct 93 14:19:47 PDT

\input phyzzx

\def\ls#1{_{\lower1.5pt\hbox{$\scriptstyle #1$}}}

\def\SCIPP{\centerline {\it Santa Cruz Institute for Particle Physics}
  \centerline{\it University of California, Santa Cruz, CA 95064}}
%\draft
\overfullrule 0pt
\Pubnum{SCIPP 93/33}
\titlepage
\date{Oct, 1993}
\pubtype{ T}     % T, E, or T/E  (Theory or Experimental)
\vskip1cm
\title{{Detecting Technibaryon Dark Matter}
\foot{Supported in part by the U.S. Department of Energy
and the Texas National Research Laboratory Commission under grant
numbers RGFY 93-263 and RGFY 93-330.}}
\author{John Bagnasco, Michael Dine, and Scott Thomas
}
%\address{}
\SCIPP
\vskip.5cm
\vbox{
\centerline{\bf Abstract}

\parskip 0pt
\parindent 25pt
\overfullrule=0pt
\baselineskip=18pt
\tolerance 3500
\endpage
\pagenumber=1
\singlespace
\bigskip

The technibaryon constitutes a possible dark matter candidate.
Such a particle with electroweak quantum numbers is already
nearly ruled out as the dominant component of the galactic dark
matter by nuclear recoil experiments. Here, the scattering of
singlet technibaryons, without electroweak quantum numbers,
is considered. For scalar technibaryons the most important
interaction is the charge radius. The scattering rates are
typically of order $10^{-4}$ (kg keV day)$^{-1}$ for a
technicolor scale of 1 TeV. For fermionic technibaryons the
most important interaction is the magnetic dipole moment.
The scattering rates in this case are considerably larger,
typically between $10^{-1}$ and 1 (kg keV day)$^{-1}$,
depending on the detector material. Rates this large may be
detectable in the next generation of nuclear recoil experiments.
Such experiments will also be sensitive to quite small
technibaryon electric dipole moments.
}

\singlespace
\bigskip

\chapter{Introduction}

\REF\srednicki{See, for example, the collection of articles in
{\it Particle Physics and Cosmology:  Dark Matter},
edited by M. Srednicki,
(North Holland, Amsterdam, 1990).}

\REF\technidm{S. Nussinov, Phys. Lett. B {\bf 165} (1985) 55;
R. Chivukula and T. Walker, Nucl. Phys. B {\bf 329} (1990) 445.}

\REF\barr{S. Barr, R. Chivukula, and E. Farhi, Phys. Lett. B
{\bf 241} (1990) 387.}

\REF\kaplan{D. Kaplan, Phys. Rev. Lett. {\bf 68} (1992) 741.
This paper discusses more general models with dark matter
candidates carrying anomalous U(1)'s.}

Technicolor and supersymmetry are the most attractive alternatives
for electroweak symmetry breaking.
Each also
provides an interesting class of dark matter candidates.
The idea that the lightest supersymmetric particle might
be the dark matter has been widely explored.  For a substantial
range of the SUSY parameter space, $\Omega \sim 1$,
where $\Omega$ is the average density divided by the critical
density.\refmark{\srednicki}
Technicolor is currently
the less popular alternative; there are many obstacles to
developing a workable model, much less an
attractive one.
However, it can naturally give rise to a suitable dark
matter candidate,  the technibaryon.\refmark{\technidm}
In most technicolor models, there is a  technibaryon symmetry
analogous to ordinary baryon number.
%violated only by
%anomalies (and possibly by higher dimension operators, arising
%from GUTS or elsewhere.)
At low temperatures the lightest
technibaryon is very nearly stable.
If technibaryon number possesses an $SU(2)_L$ anomaly, some linear
combination of baryon and technibaryon numbers will be anomaly free.
An initial net baryon number in the very early universe is then
converted to roughly equal baryon and technibaryon asymmetries by
electroweak sphaleron processes.\refmark{\barr}
Also, baryon production at the electroweak
phase transition will lead to roughly equal technibaryon
production.\refmark{\kaplan}
These mechanisms naturally give $\Omega_B / \Omega_b \sim
{\cal O}(\Lambda_{TC} / m_n)$
times a model-dependent suppression,
where $\Lambda_{TC}$ is the technicolor
scale, $m_n$ the nucleon mass, and
$\Omega_B$ and $\Omega_b$ are for the technibaryons and baryons
respectively.

\REF\hisotope{P. Smith {\it et. al.,}, Nucl. Phys. B {\bf 206}
(1982) 333;
T. Memmick {\it et. al.,} Phys. Rev. D {\bf 41} (1990) 2074;
P. Verkerk {\it et. al.,} Phys. Rev. Lett. {\bf 68} (1992) 1116.}

\REF\sastro{G. Starkman, A. Gould, R. Esmailzadeh, and
S. Dimopolous, Phys. Rev. D {\bf 41} (1990) 3594;
D. Caldwell, in {\it Proceedings of the International
School of Astroparticle Physics}, edited by D. Nanopoulos
(World Scientific, Singapore, 1991) p. 375.}

\REF\geexp{S. Ahlen, {\it et. al.,} Phys. Lett. B {\bf 195} (1987)
60; D. Caldwell, {\it et. al.}, Phys. Rev. Lett. {\bf 61} (1988)
510; D. Reusser, {\it et. al.,} Phys. Lett. B {\bf 255} (1991) 143.}

\REF\singlet{S. Nussinov, Phys. Lett. B {\bf 179} (1986) 103.}

\REF\nussinov{S. Nussinov, Phys. Lett. B, {\bf 279} (1992) 111.}

\REF\chivukula{R. Chivukula, A. Cohen, M. Luke, and M. Savage,
Phys. Lett. B {\bf 298} (1993) 380.}

If technibaryons do constitute the dark matter, the question
of direct detection arises.
Searches for anomalously heavy isotopes\refmark{\hisotope}
and astrophysical constraints\refmark{\sastro}
exclude electrically charged and strongly interacting
particles.
Experiments sensitive to nuclear recoil by elastic scattering
rule out particles with weak isospin up to a mass of a few TeV
as the dominant
component of the galactic dark matter.\refmark{\geexp}
Singlet technibaryons, without electroweak quantum numbers,
will be considerably harder to detect in recoil
experiments.\refmark{\singlet}
Most discussions of the detection of such particles have
focussed on residual color interactions.
Nussinov suggested that these might be strong enough to
allow detection.\refmark{\nussinov}  Subsequently,
Chivukula {\it et. al.}, \refmark{\chivukula}
by examining possible
effective operators which would describe such interactions,
argued that this is not the case.
However, these treatments ignored electromagnetic interactions,
associated with possible charge radii and magnetic moments.

%These turn out to be more important than any residual
%strong interactions.

In this letter a systematic analysis of the effective
operators leading to elastic scattering of
singlet technibaryons on nuclei is presented.
Since the technibaryon carries a conserved quantum number
for a continuous symmetry
(technibaryon number), the fields are either complex bosons
or Dirac fermions.
This allows additional effective operators not possible
for real bosons or Majorana fermions.
Depending on the details of the technicolor theory,
the lightest technibaryon may be either
bosonic or fermionic.\foot{Singlet fermionic technibaryons
require singlet or vector representations of
$SU(2)_L$.  We thank R. S. Chivukula for this comment.}
Each case will be considered separately.
For scalar bosonic technibaryons the most important operator turns
out to be the charge radius.
This couples coherently to the nuclear charge.
The scattering rates are quite small, however, being suppressed
by four powers of the technicolor scale.
The scattering rates are considerably larger for fermionic
technibaryons.
Here the leading operator is the magnetic moment.
Just as for QCD, technifermions are expected to have
(relatively) large magnetic moments, of the order of the Dirac
moment for a comparable charged particle.\refmark{\nussinov}
This moment couples to the nuclear magnetic moment and
coherently to the nuclear charge.
The scattering rate is suppressed by only two powers of the
technicolor scale.
If technifermions comprise the galactic dark matter,
the scattering rates are large enough to be detected in the
next generation of experiments.
The fermionic case possesses another interesting feature.
It is possible for fermionic technibaryons to have
an electric dipole moment.
Such a CP violating effect would be expected to be suppressed
by a large scale associated with CP violation, $\Lambda_{CP}$,
e.g. some
extended technicolor scale.
Even so, this moment couples coherently to the
nucleus with a large infrared enhancement.
If $\Lambda_{CP}$ is much larger than $\Lambda_{TC}$ however,
the magnetic moment still gives the dominant scattering.

Although the analysis presented here is for technibaryons,
it is generally applicable to any dark matter particle which
carries a quantum number of a continuous symmetry,
but is a singlet under
the standard model gauge groups.\refmark{\kaplan}
The effective operators and resulting scattering rates
for scalar and fermionic technibaryons are
discussed in the next two sections.
The final section addresses the question of detectability
and the possibility of experimentally
distinguishing
technibaryons from the supersymmetric neutralino.

\chapter{Scalar Technibaryons}

The coupling of a singlet technibaryon to nuclei can be regarded
as arising from effective operators defined
just above the hadronic scale,
${\cal O}$(1 GeV).
At this scale the relevant degrees of freedom are the photon,
gluon, and light quarks.
The effective operators are bilinears in the technibaryon
field multiplied by the light field operators.
The light
field operators are required to be $SU(3)_C \times U(1)_Q$
invariant and Lorentz covariant.
First consider the coupling to the photon and gluons.
At dimension three the only operator involving these gauge bosons
which can couple to a singlet scalar technibaryon  is
$$
\partial_{\mu} F^{\mu \nu}
\eqn\hyperrad
$$
where $F^{\mu \nu}$ is the electromagnetic field strength.
Using the equation of motion
$\partial_{\mu} F^{\mu \nu} = j_Q^{\nu}$,
reveals that this operator couples the technibaryon to the
electromagnetic current.
$\partial_{\mu} F^{\mu \nu}$ is therefore proportional to
the charge radius operator.
This will couple coherently to the nuclear charge.
At dimension four the potentially important
gauge boson operators are
$$
G_{a \mu \nu} G_a^{~\mu \nu}  ~~~~~
G_{a \mu \rho} G_{a ~\nu}^{~\rho}
\eqn\gg
$$
$$
G_{a \mu \nu} \tilde{G}_a^{~\mu \nu} ~~~~~
G_{a \mu \rho} \tilde{G}_{a ~\nu}^{~\rho}
\eqn\ggd
$$
where $G_{a \mu \nu}$ is the $SU(3)_C$ field strength and
$\tilde{G}_{a \mu \nu}$ its space time dual.
Analogous operators for the electromagnetic field will be
less important than the dimension three operator
\hyperrad.
The operators \gg~have the correct Lorentz structure to
couple coherently to the entire nucleus at low momentum.
However, \ggd~couple to the nuclear spin at low momentum and
will lead to scattering rates smaller
than from \gg~by at least
${\cal O}(A^2)$, where $A$ is the atomic number.

The lowest dimension operators involving the light quarks are
of the form $\bar{q} \Gamma q$, where $\Gamma$ is a Dirac matrix.
These arise (aside from \hyperrad)
{}from multigluon exchange (analogous to
Van der Waals forces) and exchanges of ETC gauge bosons.
The former make contributions to the scattering rates suppressed
by ${\cal O}(\alpha_s^2)$.
The latter are suppressed by
ratios of the technicolor and extended technicolor ($\Lambda_{ETC}$)
scales.\refmark{\chivukula}
Therefore, only the gauge boson operators \hyperrad~and
\gg, which give a coherent nuclear coupling, need be considered.

\REF\georgi{H. Georgi, Phys. Lett. B {\bf 240} (1990) 447.}

The momentum transfer in nuclear recoil is much less than the
nuclear or technibaryon mass.
The heavy field formalism is therefore useful in identifying
the effective operators
for the technibaryons.\refmark{\georgi}
In the heavy field limit
the scalar technibaryon fields, $\phi_v$, are taken
to have definite four velocity $v_{\mu}$.
These fields are related to the full field $\phi$
by\refmark{\chivukula}
$$
\phi_v(x) = \sqrt{2 M} e^{i M v_{\mu} x^{\mu} } \phi(x)
\eqn\phiv
$$
where $M$ is the technibaryon mass assumed to be of
order $\Lambda_{TC}$.
The redefinition \phiv~amounts to factoring out
of $\phi(x)$ the ``fast'' space time dependence associated with
$M v_{\mu}$ from the ``slow'' variation associated with
the momentum transfer $q_{\mu} = p_{\mu} - M v_{\mu}$.
Explicit derivatives, $\partial_{\mu}$,
acting on $\phi_v(x)$ therefore give
factors of $q_{\mu}$.
The heavy field formalism gives a convenient
parametrization of the nonrelativistic expansion.
Lorentz covariant operators are formed from bilinears of
the technibaryon fields, $\partial_{\mu}$, and $v_{\mu}$.

\REF\nda{A. Manohar and H. Georgi, Nucl. Phys. B {\bf 234} (1984) 189.}

\REF\shifman{M. Shifman, A. Vainshtein, V. Zakharov, Phys. Lett B
{\bf 78} (1978) 443.}

The coefficients of the effective operators coupling
technibaryons to the gauge fields are not
calculable from first principles since the
technibaryon is a strongly coupled bound state.
The coefficients are therefore estimated using the rules of
naive dimensional analysis.\refmark{\nda}
Up to dimension seven, the operators coupling the
techniboson to the gauge field operators \hyperrad~- \gg~are
$$
{e \over \Lambda_{TC}^2 } ~
\phi_v^* \phi_v ~ v_{\nu} ~ \partial_{\mu} F^{\mu \nu}
\eqn\crad
$$
$$
{g_s^2 \over 2 \Lambda_{TC}^3 }~
\phi_v^*  \phi_v ~ G_{a \mu \nu} G_a^{~\mu \nu}
\eqn\pola
$$
$$
{g_s^2 \over 2 \Lambda_{TC}^3 } ~
\phi^*_v  \phi_v ~v_{\mu} v_{\nu} ~G_a^{~\mu \rho} G_{a ~\rho}^{~\nu}
\eqn\polb
$$
where $\Lambda_{TC} = 4 \pi f$, and $f$ is related in a model-dependent
way to the electroweak symmetry breaking scale.
In most technicolor models, $\Lambda_{TC} \sim 1$ TeV.
As long as the techniboson constituents are charged and colored,
\crad~- \polb~should give the correct magnitude up to incalculable
coefficients of ${\cal O}(1)$.
For simplicity of notation these ${\cal O}(1)$
corrections are absorbed into $\Lambda_{TC}$ below.
\crad~gives a technibaryon charge radius of
$r_c = \sqrt{6} / \Lambda_{TC}$,
\foot{As a ``test'' of naive dimensional analysis,
applying these rules
to the neutron gives a charge radius roughly a factor of two smaller
than the experimental result.}
while
\pola~and \polb~contribute to the gluonic
polarizabilities.\refmark{\chivukula}
Even though \pola~and \polb~are suppressed by an additional
power of $\Lambda_{TC}$ relative to \crad, the relatively
large gluonic matrix elements in the nucleus make them potentially
important.
Consider for example the operator \pola.
The matrix element in the nucleus at zero momentum transfer
is related via the conformal anomaly to the
nucleus mass, $m_N$, by
$\langle N | \theta^{\mu}_{~\mu} | N \rangle = - m_N \bar{N} N$,
where
$\theta^{\mu}_{~\mu} = - {b g^2 \over 32 \pi^2}
    G_{a \mu \nu} G_{a}^{~\mu \nu}$
(ignoring the light quark contribution to $\theta^{\mu}_{~\mu}$),
and $b = 11 - (2/3) n_{\ell}$ is the $\beta$ function coefficient
for $n_{\ell}$ light flavors.\refmark{\shifman}
Using this matrix element, the cross section at small momentum
transfer can be compared to the charge radius cross section,
$$
{\sigma_{GG} \over \sigma_{\rm charge~radius} } \simeq
{ 16 \pi^2 \over b^2} \left( {A m_n \over Z \alpha \Lambda_{TC}}
      \right)^2
$$
where $m_n$ is the nucleon mass.
For $\Lambda_{TC} \simeq$ 1 TeV this ratio is $\sim$ .1.
The matrix element of \polb~can be related to the matrix
element of the gluon energy momentum tensor.
A similar ratio to the
charge radius cross section is expected.
The charge radius should therefore give the dominant coupling.

\REF\ellis{J. Ellis and R. Flores, Phys. Lett. B {\bf 263} (1991) 259;
J. Ellis and R. Flores, Phys. Lett. B {\bf 175} (1993) 175.}

In the nonrelativistic limit the differential cross section
{}from \crad~is
$$
{d \sigma \over d E_R} =
{ 16 \pi (Z \alpha)^2 m_N^2 \over
  \Lambda_{TC}^4 E_R^{\rm max} (1 + m_N / M)^2 } ~
|F_c(E_R)|^2
\eqn\sigrad
$$
where $E_R$ is the nuclear recoil kinetic energy in the lab frame,
$E_R^{\rm max} = 2 m_N V^2 / (1+m_N/ M)^2$ is the
maximum recoil energy,
%$m_N$ is the nucleus mass,
and $V$ the technibaryon velocity in the lab frame.
$F_c(E_R)$ is a form factor which accounts for the loss
of coherence over the nucleus at finite momentum transfer
($\vert F_c(0)\vert =1$).
For the numerical estimates below a Gaussian distribution
of nuclear charge is assumed,
$|F_c(E_R)|^2 = e^{- {2 \over 3} m_N R^2_c E_R}$,
with an rms radius
$R_c = (.3 + .89 ~A^{1/3})$ fm.\refmark{\ellis}
For Ge with a recoil energy of 10 keV (near the threshold energy
of the current generation of experiments)
$|F_c(E_R)|^2 \simeq .83$.
The differential scattering rate per unit detector mass, $m$,  is
given by
$$
{d R \over dm~dE_R} = { \rho \over m_N ~M} ~
\langle {d \sigma \over dE_R} V \rangle
\eqn\rate
$$
where $\rho$ is the technibaryon density, and
$$
\langle {d \sigma \over dE_R} V \rangle =
\int dV~f(V) ~V~ {d \sigma \over dE_R}(V,E_R)
$$
$$
\equiv V_0 {d \sigma \over dE_R}(V_0,E_R)
\eqn\vnot
$$
where $f(V)$ is the technibaryon velocity distribution
in the lab frame,
and $V_0$ is a suitably weighted average velocity.
The rates from \sigrad~are quite small,
being suppressed by four powers of $\Lambda_{TC}$.
Numerically, for Ge
$$
{dR \over dm~dE_R} \simeq 2 \times 10^{-4}
 ~|F_c(E_R)|^2~
 \Lambda_{\rm TeV}^{-4}~ M_{\rm TeV}^{-1}~ \rho_{.3}~
  V_{320}^{-1}
{}~({\rm kg~keV~day})^{-1}
$$
where $\Lambda_{\rm TeV} \equiv \Lambda_{TC} / {\rm TeV}$,
$M_{\rm TeV} \equiv M / {\rm TeV}$,
$\rho_{.3} \equiv \rho /(.3~ {\rm GeV ~cm}^{-3})$, and
$V_{320} \equiv V_0 / (320~ {\rm km~ s}^{-1})$.
While this is substantially larger than the rates implied in Ref.
[\chivukula], it is still well below
the current bound for Ge of
$\sim 3$/(kg keV day)$^{-1}$.\refmark{\geexp}

\chapter{Fermionic Technibaryons}

\REF\jenkins{E. Jenkins and A. Manohar, Phys. Lett. B {\bf 255}
(1991) 558.}

\REF\magnino{Aspects of the low energy magnetic dipole
moment scattering have been considered in a different context
by S. Raby and G. West,
Phys. Lett. B {\bf 200} (1988) 547.}

In addition to the operator \hyperrad, technibaryons with spin can
couple to the dimension two electromagnetic field strength,
$F^{\mu \nu}$, via dipole moments.
In virtually any conceivable
technicolor scenario, some of the technibaryon constituents
will be charged, and such couplings will arise.
This leads to much more dramatic effects.

For definiteness spin ${1 \over 2}$ technibaryons
will be considered.
The extension to higher spin is straightforward.
Again, the heavy field formalism is useful for writing the effective
operators.
The technifermion field of definite four velocity,
$\Psi_v$, is related to the full field $\Psi$ by\refmark{\georgi}
$$
\Psi_v(x) = e^{i M {\not  v} \cdot x} \Psi(x)
\eqn\hfermion
$$
At dimension five the only operator suppressed by a single
power of the technicolor scale is the magnetic moment
$$
  {e \over \Lambda_{TC} } ~
  \epsilon_{\mu \nu \rho \sigma} v^{\rho}
    \bar \Psi_v S^{\sigma} \Psi ~F^{\mu \nu}
\eqn\magmoment
$$
where $S^{\mu}$ is the spin operator.\refmark{\georgi,\jenkins}
The coefficient is taken from the rules of naive dimensional
analysis.
This gives the technifermion a magnetic moment
$\mu = e / \Lambda_{TC}$.

\REF\nextdm{B. Young, {\it et. al.}, in {\it Proceedings of the
XXVI International Conference on High Energy Physics}, edited by
J. Sanford (American Institute of Physics, New York, 1993)
p. 1260.}

The technifermion magnetic moment couples to the nuclear
magnetic moment and coherently to the current produced
by the nuclear charge (in the technifermion
rest frame).\refmark{\magnino}
The differential cross section
in the nonrelativistic limit is
$$
{d \sigma \over dE_R} =
 {4 \pi \alpha^2 \over \Lambda_{TC}^2 (1+m_N/M)^2 E_R^{\rm max} }
 \left\{ { 2(J+1) \over 3 J}
         \left( {\mu A \over \mu_n} \right)^2
          |F_s(E_R)|^2
 \right.
$$
$$
\left.
+~ Z^2 \left( (1+m_N/M)^2 {E_R^{\rm max} \over E_R}
              - {2 m_N \over M} - 1  \right)
                |F_c(E_R)|^2
\right\}
\eqn\magsig
$$
where $J$  is the nuclear spin, $\mu / \mu_n$ is the
nuclear magnetic moment in units of the nuclear Bohr magneton
($ = e / 2 m_n$),
%$A$ is the atomic number,
and
$F_s(E_R)$ is a form factor for the nuclear spin.
For the numerical estimates below the somewhat
crude assumption is made of a Gaussian distribution of spin within
the nucleus with spin radius $R_s = 1.25~R_c$.\refmark{\ellis}
For large recoil energy the nuclear magnetic moment and
coherent contributions are typically the same order.
The coherent contribution has an infrared singularity,
giving a small enhancement for experimentally
accessible recoil energies.
For Ge in natural abundance, the coherent nuclear coupling
dominates.
Numerically, for $E_R << E_R^{\rm max}$,
$$
{dR \over dm~dE_R} \simeq 1 \times 10^{-2}~
   {E_R^{\rm max} \over E_R} ~
   |F_c(E_R)|^2 ~ \Lambda_{\rm TeV}^{-2} ~M_{\rm TeV}^{-1}~
      \rho_{.3} ~ V_{320}^{-1} ~({\rm kg~keV~day})^{-1}
$$
For $E_R = 10$ keV, $dR /( dm~dE_R) \simeq .15
{}~\Lambda_{\rm TeV}^{-2} ~ M_{\rm TeV}^{-1} ~ \rho_{.3} ~
    V_{320}^{-1}~ $(kg keV day)$^{-1}$.
Scattering rates of this order may be accessible to the next
generation of Ge experiments.\refmark{\nextdm}

The coupling to the nuclear magnetic moment can be significant
in elements with large magnetic moments.
For $^{93}$Nb with $J={9 \over 2}$ and
$\mu \simeq 6.17~\mu_n$, the
magnetic moment gives the dominant contribution even for
$E_R = 10$ keV.
Numerically,
$$
{dR \over dm~dE_R} \simeq ~ 2 ~|F_s(E_R)|^2
   ~ \Lambda_{\rm TeV}^{-2} ~M_{\rm TeV}^{-1}~
      \rho_{.3} ~ V_{320}^{-1} ~({\rm kg~keV~day})^{-1}
$$
This is comparable to the scattering rate for a technibaryon
with weak isospin.
The large magnetic moment makes Nb a good candidate material
to search for dark matter technifermions.
The nuclear magnetic moment scattering is also significant for
NaI and CaF$_2$.

In addition to the magnetic dipole moment \magmoment, a technifermion
can possess an electric dipole moment.
This P and T violating operator should be suppressed by the scale
associated with CP violation, $\Lambda_{CP}$.
In a realistic model this is presumably related to an
extended technicolor scale responsible for
light fermion masses.
If the lowest dimension P and T  violating operators
among the technibaryon constituents are four-Fermi type
of dimension six, the electric dipole moment must be suppressed
by two powers of $\Lambda_{CP}$.
Including this suppression then gives the estimate
$$
{e \over \Lambda_{TC} }
  \left({ \Lambda_{TC} \over \Lambda_{CP} } \right)^2 ~
 v_{\mu}~ 2 \bar \Psi_v  S_{\nu} \Psi_v ~
      F^{\mu \nu}.
\eqn\edm
$$
where the coefficient of the CP violating dimension six
operator is absorbed in $\Lambda_{CP}$.
This gives the technifermion an electric dipole moment
$d = e \Lambda_{TC} / \Lambda_{CP}^2$.
In the nonrelativistic limit, the differential cross
section is
$$
{d \sigma \over d E_R} =
  {8 \pi (Z \alpha)^2 \over
   (\Lambda_{CP}^2 / \Lambda_{TC})^2  (1 + m_N/ M )^2
     E_R^{\rm max} }
  {m_N \over E_R}
   |F_c(E_R)|^2
\eqn\sigedm
$$
The ratio of
electric dipole scattering to the coherent part
of the magnetic dipole scattering is
${\cal O}( m_N \Lambda_{TC}^4 / E_R^{\rm max} \Lambda_{CP}^4)$.
The enhancement occurs because the electric dipole couples
directly to the nuclear charge rather than the
nuclear current.
Roughly, for $\Lambda_{CP} < 30 ~\Lambda_{TC}$
the electric dipole is more important.
Numerically, for Ge
$$
{dR \over dm~dE_R} \simeq 1 \times 10^4
{E_R^{\rm max} \over E_R}
\left( {\Lambda_{TC} \over\Lambda_{CP} } \right)^4
{}~|F_c(E_R)|^2
{}~ \Lambda_{\rm TeV}^{-2}
{}~M_{\rm TeV}^{-1} ~\rho_{.3} ~V_{320}^{-1}
{}~({\rm kg~keV~day})^{-1}
$$
With $E_R$ = 10 keV, the current bound from Ge already
gives a limit of
$\Lambda_{CP} > 14 ~( \Lambda_{\rm Tev}^2 \rho_{.3} /
 M_{\rm TeV} V_{320} )^{1/4}$ TeV.

\chapter{Conclusions}

\REF\susy{M. Goodman and E. Witten, Phys. Rev. D {\bf 31} (1985) 3059;
K. Griest, Phys. Rev. D {\bf 38} (1988) 2357.  For recent calculations
including constraints from $\Omega_{\rm SUSY}$ see
A. Bottino, {\it et. al.}, Phys. Lett. B {\bf 295} (1992) 330;
A. Bottino, {\it et. al.}, preprint DFTT 38/93;
M. Drees and M Nojiri, preprint MAD/PH/768.}

The detectability of singlet
dark matter technibaryons depends sensitively
on the spin.
For scalar technibaryons the dominant coupling is through the
charge radius.
The rate from this coupling is suppressed by four powers of
the technicolor scale.
The total scattering rate for Ge, assuming a threshold recoil
energy of 5 keV, is $\sim 1 \times 10^{-2}$ (kg day)$^{-1}$
for $\Lambda_{TC} \simeq$ 1 TeV.
While this is well beyond the reach of current experiments, it is
comparable to spin dependent scattering rates for the
neutralino of
supersymmetry.\refmark{\ellis,\susy}
For fermionic technibaryons the dominant coupling is through the
magnetic moment.
The scattering rate is suppressed by two powers of the technicolor
scale.
The total rate can be sizeable.
For Ge in natural abundance, assuming a threshold of 5 keV,
the rate is $ \sim 3$ (kg day)$^{-1}$ for
$\Lambda_{TC} \simeq$ 1 TeV.
This is comparable to the spin independent rates for the
neutralino.\refmark{\ellis,\susy}
Even larger rates are attainable in elements with sizeable nuclear
magnetic moments.
For Nb, again assuming a threshold energy of 5 keV, the total
rate is $\sim 35$ (kg day)$^{-1}$.
In principle, a technibaryon electric dipole moment could also
lead to sizeable rates.
However, in a ``realistic'' model with
$\Lambda_{CP} \sim \Lambda_{ETC} \sim 10^3~ \Lambda_{TC}$,
the magnetic moment still dominates the scattering.

An important feature of singlet technibaryon scattering is that the
dominant couplings are electromagnetic.
This is in contrast to the spin independent coupling of the neutralino
which is typically dominated by operators such as \gg.
In principle then, comparison of scattering rates in different
materials could allow the technibaryon to be distinguished from
the neutralino.
In addition, the coherent part of the magnetic moment scattering
has a distinctive energy dependence at low energy (Eq. \magsig).
Finally, it is worth noting that the analysis presented here
is generally applicable to any dark matter particle carrying a
quantum number for a continuous symmetry,
but which is a singlet under the standard model
gauge groups.\refmark{\kaplan}

\refout

\end